\documentstyle[12pt,aaspp]{article}
\def\etal{{\it et al.~}}
\def\kms{km~s$^{-1}$~}

\def\W50{W$_{50}$~}
\journalid{337}{Someday 1999}
\articleid{11}{14}
\begin{document}
\title{HI 2334+26: An Extended HI Cloud near Abell 2634}

\author{Riccardo Giovanelli, Marco Scodeggio, Jos\'e M. Solanes, Martha P.
Haynes, H\'ector Arce}
\vspace{.15in}
\affil{Center for Radiophysics and Space Research \\
and National Astronomy and Ionosphere Center\altaffilmark{1},\\
Cornell University, Ithaca, NY 14853}
\and
\author {Shoko Sakai}
\vspace{.15in}
\affil{Dept. of Physics and Astronomy, Dartmouth College, Hanover,
NH 03755}

\altaffiltext{1}{The National Astronomy and Ionosphere Center is operated by
Cornell University under a cooperative agreement with the National Science
Foundation.}
\hsize 6.2 truein

\begin{abstract}
We report the serendipitous discovery of a large HI cloud with an
associated HI mass of $6(\pm1.5)\times 10^9 h^{-2}$
M$_\odot$ and a heliocentric velocity 8800 \kms, located near the periphery
of the cluster of galaxies Abell 2634. Its velocity field
appears to be very quiescent, as no gradients in the peak velocity
are seen over its extent of 143$ h^{-1}$ by 103$ h^{-1}$ kpc. The
distribution of gas is poorly resolved spatially, and it is thus
difficult at this time to ascertain the nature of the cloud. At least
two relatively small, actively star--forming galaxies appear to be
embedded in the HI gas, which may (a) be an extended gaseous envelope
surrounding one or both galaxies, (b) have been spread over a large
region by a severe episode of tidal disruption or (c) have been affected
by the ram pressure resulting from its motion through the intracluster
gas of A2634.

\end{abstract}


\pagebreak

\section{Introduction}

Atomic hydrogen in a galaxy is a convenient
tracer of the potential for future star formation and a probe of the
kinematical field.  Because normal spiral galaxies
contain an amount of HI predicted by their optical radii
(see reviews by Giovanelli $\&$ Haynes 1987; Roberts $\&$ Haynes 1994, and
refs. therein), the neutral hydrogen content of a spiral galaxy is a suggestive
indicator of  disturbances in the galaxy's evolutionary history. Collisions,
tides and intracluster interactions can remove gas, lowering a disk galaxy's
HI content or possibly create a  star--forming disk in a system previously
without one. In the case of tidal interactions and gas sweeping, the HI
 distribution can also provide a glimpse of the disruptive event as the gas,
removed from its normal disk location, is flung to large distances.
Numerous examples (see refs. above) exist of HI displaced from the parent
galaxy by such processes. In many instances the mapping of the HI distribution
 provides the clues needed to constrain simulations of the disruptive event
 (e.g. Stanford $\&$ Balcells 1991; Smith 1994).

Extended HI distributions may also indicate the relative youth of systems
still in the process of collapse. In I Zw 18, whose youth is reflected in its
extreme low metal abundance, the compact starbursting components
are embedded in a much larger HI envelope (Viallefond \etal 1987;
Lequeux $\&$ Viallefond 1980). HI 1225+01 is formed by two main HI
condensations, one of which is associated with an actively star forming blue
dwarf galaxy. The other condensation emits no detectable starlight
(Giovanelli \etal 1991). It can be speculated that the optically visible
component has only recently breached the star formation threshold
(Chengalur \etal 1994). Taylor \etal (1993) further hypothesize that
the star formation events in blue compact dwarfs are triggered
by interactions with nearby HI companions.

Although less clearly traceable,
the link between extended HI distributions, star formation
and the formation of disks has also been discussed in scenarios
of severe tidal disruption and merging.  Some active
galaxies are seen to be embedded within a more
extensive HI distribution (see for example Heckman \etal 1982).
Appleton \etal (1990) show evidence of the formation of
a disk parallel to the minor axis within the elliptical
NGC 5903 as the result of recent accretion. Barnes $\&$ Hernquist
(1992) have suggested that dwarf galaxies may form in the tidal debris
as possibly indicated in such systems as the Antennae (Mirabel \etal 1991).
In compact groups, HI is often seen displaced from the member galaxies
(Shostak \etal 1984; Williams $\&$ van Gorkom 1988).
Williams \etal (1991) suggest even further that the HI distribution
in compact groups may be a good indicator of the evolutionary stage of the
group itself.

Giovanelli $\&$ Haynes (1989) have suggested
that HI 1225+01, a system with a very large HI--to--optical
size ratio, has remained intact due to its relative isolation.
Compact groups and clusters, however, present a harsh environment
to the thermally fragile HI gas through
tidal and dynamical stirring, as well as contact with the hot intracluster
medium. While the interpretation of giant HI envelopes may be ambiguous,
it is thus surprising to find weakly bound HI clouds of large extent
in high density environments.

Here, we report the serendipitous discovery of an extended
HI cloud, of unusually large extent, located near the periphery
of the cluster of galaxies Abell 2634.  Observations of the 21 cm
HI line emission and follow--up optical imaging and spectroscopy
are presented in Section 2. Although the current HI observations
are of inadequate spatial resolution to allow an unambiguous
interpretation of the nature and origin of the HI cloud, we
briefly discuss in Section 3 possible interpretative outlines.
Intrinsic properties are computed for $H_0 = 100 h$ \kms Mpc$^{-1}$.

\section{Observations}

In the course of a 21 cm line survey of galaxies in the cluster A2634
undertaken with the 305m telescope of the Arecibo Observatory
(Scodeggio \etal 1995, hereinafter SSGH; Giovanelli \etal, in preparation),
we have obtained HI line spectra for a number of previously uncatalogued
galaxies identified by us from a visual examination of the
Palomar Observatory Sky Survey (POSS) prints to lie in or near the cluster.
One such object is a spiral galaxy at RA(1950) = 23$\rm ^h$ 33$\rm ^m$
57$\rm ^s$, Dec(1950) = 26$^\circ$ 22.7' --- hereinafter referred to as
galaxy A ---  observed and detected in the HI line in July of 1993.

The HI emission spectrum of galaxy A is shown in Figure 1a.
In addition to an emission feature identified with the optical galaxy at
a systemic heliocentric velocity of $9212\pm 4$ \kms, the spectrum
exhibits a feature of comparable flux near V$_{hel}$=8800 \kms.
When a nearby galaxy at RA(1950) = 23$\rm ^h$34$\rm ^m$16$\rm ^s$,
Dec(1950) = 26$^\circ$23.1\arcmin --- hereinafter referred to as
galaxy B --- was observed, the same feature at 8800 \kms was detected
again, as illustrated in Figure 1b. The relative intensity at
the two positions suggested that the 8800 \kms feature was extended,
and a coarse mapping of the region ensued.

Figure 2 summarizes the results of HI spectra taken at 45 positions
on a mapping grid with the Arecibo dual circular 40--foot tunable 22 cm feed.
This feed has a half-power beamwidth of 3.3\arcmin.  Most spectra had an
on--line spectral resolution of 8 \kms, degraded to 16 \kms
after smoothing. The typical rms noise level of each smoothed spectrum
is 0.85 mJy. A few observations near the center of the apparent HI distribution
were made with a resolution twice as high, e.g. figure 1b.
In Figure 2, the locations of observed spectra are
indicated by either crosses (no detection) or filled circles (detections);
the area within each filled circle is proportional to the integrated flux of
the 8800 \kms feature at each position. Observations coincident with
galaxies A and B are labeled, as is that of a third galaxy (``C'') to be
discussed later. The peak velocity at each position at which HI is detected
is indicated under each symbol; associated velocity errors, which are dependent
on signal--to--noise ratio, vary between 3 and 20 \kms, with the larger
errors near the cloud periphery.
The half--power beam size of the 22 cm feed is illustrated in the
figure to emphasize the limited spatial resolution of this set
of observations.  Because the 22 cm feed is also characterized
by the presence of a first sidelobe ring of 5.6\arcmin\ in diameter, peaking
at about the 10\% level of the main beam, caution is necessary in
the interpretation of extended emission. Detections are identified
only where flux densities are well in excess of what would
be expected from sidelobe contamination; by comparison, galaxy A is only
detected when the beam is on source or on points less than two thirds
of a beamwidth away, as indicated by the square symbols in fig. 2. Contour
lines at equal levels of integrated flux per beam are superimposed.
In spite of the poor resolution, the HI feature is definitely extended.
Approximate deconvolution of the beam smearing effects (see Giovanelli \etal
1991 for details) suggests that
the HI extends over a region of size about 5.5\arcmin\ by 4\arcmin, roughly
the outline of the outermost solid contour. The most striking aspect
of Figure 2 is the near constancy of the velocity of the HI emission
associated with the extended HI. Despite the limitations in spatial
resolution of the current map, the continuous but quiescent nature
of the velocity field suggests that the extended HI distribution may be
a single diffuse structure seen close to face--on. In the following,
therefore, we refer to the extended HI distribution as ``the HI cloud''.

In the course of a direct imaging run made in poor
photometric conditions, two images centered on the cloud position (1800 s
exposure at B and 600 s exposure at I band) were obtained in September
1993 with the 1.3m telescope at the MDM observatory, equipped with a Loral
$2048\times 2048$ CCD chip. On--chip rebinning by a factor of two
yields images with a scale of
0.64\arcsec pixel$^{-1}$, covering a field of 10.9\arcmin\ on a side.
As shown in Figure 3, the blue image of the field contains several galaxies
with angular sizes between 10\arcsec\ and 20\arcsec; the sky density of such
objects is quite high and is related both to the nearness to the center
of A2634 and to the presence of a background cluster (see SSGH).

Galaxies with identification labels in Figure 3 were among the
targets of spectroscopic observations in the A2634 region obtained using
the 2.4m telescope at the MDM observatory in September 1993.
The spectrograph was equipped with a Loral $2048\times 2048$ CCD chip
and a low dispersion grating (300 lines mm$^{-1}$), giving spectral coverage of
 the region between 4500 and 8000 \AA. Spectra with  signal--to--noise ratios
adequate for redshift measurements were obtained for 38 galaxies, as listed
in Table 1. Reduction followed standard procedure using IRAF\footnote{IRAF is
distributed by the National
Optical Astronomy Observatories which is operated by the Association of
Universities for Research in Astronomy, Inc. under cooperative
agreement with the National Science Foundation.}.
tasks. Each frame was overscan- and
bias-subtracted, then flat-fielded using appropriate dome flats. Wavelength
calibration was performed using HeNeAr comparison spectra. Heliocentric
velocities were computed in two different ways, depending on the emission
or absorption line nature of the spectrum: using the IRAF task {\it fxcor},
that is based on the
cross-correlation algorithm described by Tonry \& Davis (1979), for E and S0
galaxies; and measuring the central wavelength of the gaussian fit to the
emission lines of spiral galaxies (typically between 3 and 5 lines were
measured for each galaxy).  Three different K giant stars were used as
templates for the cross-correlation.

Radial velocities have been obtained for several galaxies in Table 1 also
by Pinkney \etal (1993). The redshifts given for the
three galaxies A, B and C by those authors differ significantly
from the ones reported here; Pinkney \etal obtain for galaxies A, B and C
respectively 12205, 9203 and 6759 \kms. It should be noted that
the uncertainty of these former measurements is rather large, because of
limited signal to noise ratio in their spectra (see Pinkney \etal for a
 discussion of their class C2 redshift determination, and SSGH for a
broader comparison of our data and those of Pinkney \etal), and because
their spectral coverage did not extend far enough to the red to
include the H$\alpha$ line. Note that our redshift for galaxy A is obtained
from the HI spectrum shown in Figure 1a, while the two other galaxies
show such strong H$\alpha$ that the optical redshift we derive is certain.

Among the galaxies projected on the HI cloud, those identified as A,B,C,E and
G have velocities that make them likely members of A2634. Indeed, the HI
emission found at the position
of galaxies B and C cannot be separated into components associated
with the galaxy disks and the extended cloud. Three other galaxies with
measured velocities near 35,000 \kms are probably member of a background
cluster (Pinkney \etal 1993; SSGH). There are several other
faint galaxies projected on the region of HI emission, but unfortunately
no radial velocities are available for these objects.

\section{Discussion}

\subsection{System Parameters}

As discussed above, we interpret the extended HI emission as a single
coherent structure subject to the caveat of the limitations of the current
HI map. Because of its position and velocity, it is reasonable to assume
that the HI cloud is bound to the cluster A2634, but likewise, there is some
ambiguity in interpreting the redshift of the HI cloud directly in terms
of its distance. The cluster center, RA(1950) = 23$^h$35$^m$55$^s$,
Dec(1950) = 26$^\circ$44\arcmin19\arcsec, lies about half a degree to
the northeast of the cloud center.
For A2634, SSGH obtain a heliocentric systemic
velocity of 9240 \kms and a velocity dispersion of 661 \kms.
Therefore, any object at the position of the cloud with a velocity between
8000 and 10,500 \kms is very likely a cluster member (see SSGH for a detailed
discussion on cluster membership).
Assuming that A2634 is at rest with respect to the
reference frame of the cosmic microwave background radiation,
the HI cloud would have a distance of $89\,h^{-1}\rm\,Mpc$.
Alternatively, it could be located at its redshift distance of
$84\,h^{-1}$ Mpc in the foreground of the cluster, or it could be a background
 object with a significant infall velocity.
In the following we derive sizes, masses and luminosities assuming the
HI cloud and the galaxies most closely associated with it are at the cluster
distance.

The rough extent of the cloud is $143h^{-1}$ by $103\,h^{-1}$
kpc; the HI mass is $M_{\rm HI} = 6(\pm1.5)\times 10^9\,h^{-2}$ M$_\odot$.
Most noticeable is the constancy in the mean velocity of the HI and the
 narrowness of the HI line which, measured at the half--power
points, is also seen to be remarkably constant about 140 \kms.
It is to be emphasized that while the low spatial resolution of the beam
is expected to smear any existing velocity gradients, a
systematic difference in the mean velocity between opposite sides of
the HI at a level greater than 20 \kms would be easily
detected by these HI observations. For example, the receding and
approaching sides of galaxies significantly smaller than the beam
are easily identified by applying half beam pointing offsets.

The properties of galaxies A,B and C are summarized in Table 2.
Because the photometric quality of the MDM images is poor, we are
unable to assign absolute values to magnitudes and colors of the
objects in the field. We estimate the blue total magnitude of galaxy A
to be on the order of $16.3\pm0.2$, on the basis of visual inspection
of the Palomar Observatory Sky Survey plates and of digitized images
of the ``Palomar Quick Survey'' obtained from the Space Telescope Science
Institute Archives, kindly assisted by Dr. D. Golombek.
Galaxy B is about 0.65 $B$ mag fainter and galaxy C 0.70 mag
fainter than galaxy A; galaxy C also appears to be 0.2 mag bluer
than A and B (in $B-I$). Isophotal sizes of the 3 galaxies at the approximate
level of $\mu_B \sim 26$ mag arcsec$^{-2}$ are also estimated; these
values are listed in Table 2. For $h=1$, the distance modulus of A2634 is
34.7; therefore these galaxies are about 3 mag fainter than the knee of the
luminosity function M$_*$.

The velocity separation between galaxy A and the HI cloud at 8800 \kms is
large ($> 400$ \kms) and it is therefore likely that no relation exists
between the two. The other two galaxies  --  B and C -- have optical radial
velocities undistinguishable from that of the HI cloud and lie within its
boundaries. For objects bound to a cluster, coincidence in projected
position and velocity is not necessarily equivalent to spatial proximity.
However, both objects exhibit unusually
strong emission lines in their optical spectra.  In Table 2, the
equivalent widths of H$\alpha +$ [NII] are significantly larger than those
expected for normal spiral galaxies (Kennicutt \& Kent 1983).
This implies strong star formation activity, which
requires large supplies of gas. The association of galaxies B and C with the
HI is therefore likely. It is relevant to note
the association of extended HI clouds with regions of active star formation
bursts in diverse objects like I Zw 18 (Searle $\&$ Sargent 1972) and
HCG 18 (Williams $\&$ McMahon 1988).

The extent of the HI is large in comparison
with the sizes of the optical disks or even with the projected
separation of galaxies B and C on the plane of the sky, which is only
$55\,h^{-1}$ kpc. The $M_{\rm HI}/L_B$ ratio, even when we add the
luminosity contribution of both galaxies, yields a value of 1.6.
Typical values for field galaxies range between
0.2 and 0.5, for spirals (Roberts $\&$ Haynes 1994); a value of 1.6 is found
only in the upper quartile of the latest types, Sm/Im.
Other systems contain comparable amounts of HI distributed
across similar extents.  The quiescent cloud in HCG 18
has a diameter of 37.5 h$^{-1}$ kpc and an HI mass
of 6 $\times$ 10$^9~h^{-2}$M$_{\odot}$ (Williams $\&$ van Gorkom
1988). For comparison, HI 1225+01 contains 4.9  $\times$
10$^9~h^{-2}$M$_{\odot}$ of HI over an extent of $\sim$ 215 h$^{-1}$ kpc
(Giovanelli \etal 1991). Comparatively, the size and mass of the HI
cloud in Abell 2634 are high, though not uniquely so.

\subsection{Interpretative Scenarios}

We consider three possible interpretations of the origin of the HI cloud.
First, we assume that B and C are two gas--rich galaxies
currently  undergoing an encounter that results in tidal fireworks
of the type described by Toomre $\&$ Toomre (1972). Tides and
collisions are the suggested cause of the offset HI distributions
seen in Stephan's Quintet (Allen $\&$ Sullivan 1980; Shostak
\etal 1984). In a second scenario, the two galaxies are embedded in a
common gaseous envelope of the
type seen in the giant Virgo cloud HI 1225+01 (Giovanelli \etal
1991; Chengalur \etal 1994). Such a cocoon scenario has also been suggested
for the interpretation of HCG 18 as a single amorphous galaxy
(Williams $\&$ van Gorkom 1988). In a third, the HI associated
with either galaxy B, C or both, is recently swept by the ram pressure of
the intracluster medium in A2634 not unlike the case of NGC 1961
(Shostak \etal 1982).

All interpretations have to contend with the lack of Doppler
gradients in the gas, as if the relative motion
of the two galaxies and the internal motions within each
occurred largely in the plane of the sky.  Such a velocity field
is reminiscent of the HI cloud in HCG 18 (Williams $\&$ van Gorkom 1988)
but quite unlike that seen in Stephan's Quintet (Shostak \etal 1984)
or Seyfert's Sextet (Williams \etal 1991).

While we cannot separate the cloud and galaxy emission in the direction
of objects B and C, we can nonetheless ask whether it is plausible that
the galaxies are embedded in the cloud by comparing their expected
velocity spread with that observed in the HI emission at those locations.
We can crudely estimate the magnitude of internal motions in the two
galaxies using the luminosity--HI line width relation. For example,
for galaxy A, the observed
width of 182 \kms converts to about 275 \kms after applying a correction
for inclination; this value and that of the magnitude in Table 2
agree well with the luminosity--HI line width relation in
Haynes $\&$ Giovanelli (1984). Using the same relation, one should expect
widths of about 225 \kms for galaxies B and C which, after correction for
inclination should be observed between 160 and 190 \kms. While somewhat higher
 than the observed HI widths, the expected values are not untenably high, and
therefore internal motions in the galaxies may occur within the
range of velocities of the observed line widths. On
the edges of the HI cloud, however, one should expect the
gas to have a lower velocity dispersion than in the region
occupied by the galaxies. Figure 4 displays an east--west cut
of the velocity field, which crosses over the position of galaxy
C. While emphasizing again the constancy of the
mean velocity in the field, it also hints at a broadening of the HI line
widths near RA = 23$\rm^h$34$\rm ^m$10$\rm ^s$, the location of galaxy C,
and to the east.

In the cocoon scenario, the fact that
neither of the two galaxies (especially galaxy B) lies in the
region of highest column density for the gas must be explained. One may
argue that only galaxy C is associated with the HI gas and somewhat force
the issue by attributing the (smaller) offset of C from
the center of the gas to poor telescope pointing (which cannot be totally
excluded, although observations were made over a wide range of
altazimuthal configurations and a systematic pointing offset would
not likely be present in the map), and assume that B is an
interloper seen closeby only in projection. The strong emission
lines in the spectrum of B however suggest the presence of a fresh
supply of gas.

The tidal scenario is likewise plausible, but the circumstance
of a ``flat'' velocity field to a couple of tens of \kms or less
requires a very special geometry for the encounter. This model
can handle better the offset of the galaxy positions from the HI
peak, if most of the gas is visualized as initially associated with
one of the two galaxies (possibly C). Given the
small separation of the two galaxies, tidal damage can be significant,
especially if the gas was loosely connected
to the parent galaxy (e.g., in an extended disk) to start with.
Since the center of the HI cloud appears offset from both galaxies, the
interaction would have been of an exceptionally disruptive nature, although
no morphological signature of tidal distorsion is evident in the CCD image.

It is relevant to underscore that the space density of galaxies in this region
 is significantly enhanced, and that all galaxies
between 8000 and 10,500 \kms in the region are well within the
most conservative set of caustics for the cluster A2634 (SSGH). Within a
20\arcmin\ radius from galaxy C, two kinematically
separate groups of galaxies are identifiable: 9 objects with velocities
between 8300 and 9000 \kms, and 8 with velocities between 9200 and
9700 \kms, with a clear gap in between. Both groups are spiral rich,
and are presumably falling for the first time toward the cluster
center, where the loosely bound HI material is unlikely to retain
any association with individual galaxies, whether it is now in
the form of an extended envelope or of tidal appendages.
This dynamical circumstance and the symmetry of the HI distribution,
vis--a--vis the direction to the cluster core, suggest the possibility
that the positional offset of the HI cloud and the galaxy or galaxies
initially associated with it may result from the interaction of the cold HI
with the intracluster medium associated with A2634. The ram pressure stripping
 model of Gunn \& Gott (1972; eq. [62]) provides an estimate of the
expected efficiency of ram pressure ablation on galactic disks.
{}From the Eilek \etal (1984) hydrostatic equilibrium model of the hot gas
distribution in A2634 (for an X-ray map see fig. 17 of SSGH),
we infer a mass density of gas at the position of galaxy C of $2.2\times
10^{-29}h^{1/2}\rm\,g\,cm^{-3}$, while the mean disk restoring force of
a typical spiral is $1.0\times 10^{-11}\rm\,dyn\,cm^{-2}$. The ratio of
these two quantities implies that, for ram pressure stripping to be
effective, the component of the velocity of the galaxy C perpendicular
to its disk should be $\sim 100$ times larger than its observed
velocity offset (424 \kms in the reference frame of the cluster),
or that the gravitational force binding the gas to the
galaxy should have a value $\sim 100$ times smaller than the one
adopted here.  As in the case of NGC 1961 (Shostak \etal 1982;
Gottesman \etal 1983), the possibility of stripping is not
easily explained but cannot be ruled out.

Whichever the current configuration of the
gas may be, this system appears destined for further dramatic changes.
Planned synthesis observations should soon help to clarify the
picture, when further speculations will be better justified.

\acknowledgements
This research is partly based on observations carried out at the MDM
Observatory, jointly operated under a cooperative agreement between the
University of Michigan, Dartmouth College and the Massachussets Institute of
Technology.
Partial financial support was provided by NSF grants AST91--15459 to RG and
AST90--23450 and AST92--18038 to MPH. JMS also acknowledges support
by the United States--Spanish Joint Committee
for Cultural and Educational Cooperation and the Direcci\'on General de
Investigaci\'on Cient\'{\i}fica y T\'ecnica through Postdoctoral Research
Fellowships.

\vfill
\eject

{\bf Figure 1:}  \hskip 7pt   (a) HI profile taken at the position of
galaxy A, with a channel separation of 8 \kms. The emission of the
galaxy is the feature at $V_{hel} = 9212$ \kms, while that of the extended HI
 is centered near $V_{hel} = 8800$ \kms. The downgoing
feature near 8500 \kms is poorly subtracted man--made interference
(GPS satellite system). (b) HI line profile taken at the location of the peak
column density of the extended HI source, with a channel separation of 4 \kms.

{\bf Figure 2:} \hskip 7pt Map of the HI emission. Solid symbols
identify detections of HI emission; the area of each circle is
proportional to the integrated flux density within the beam, with
the largest observed value being 1.89 Jy \kms. The mean velocity
between the half--intensity
points, is displayed under each symbol with well measured emission.
Small crosses identify positions where no detection was obtained,
at a level above 0.25 Jy \kms. Solid contour intervals correspond to
levels of 1.2, 0.9 and 0.6 Jy \kms per beam, while the dashed one
is at 0.35 Jy \kms per beam. Unfilled squares identify positions
at which detection of the emission associated with galaxy A near
9200 \kms is seen; the area of each symbol is again proportional to
the integrated flux density; note that the detectability of galaxy A
falls off rapidly from the central position.
The position of galaxies A, B and C is identified.

{\bf Figure 3:} \hskip 7pt $B$ band image obtained at MDM Observatory,
flat fielded but photometrically uncalibrated. Scale is 0.64'' per
pixel. Galaxies with known redshifts, as listed in Table 1, are labeled.

{\bf Figure 4:} \hskip 7pt Cut through the velocity field of the
HI gas, at the indicated constant value of the declination, slightly
to the South of galaxy C, which is at RA = 23$\rm ^h$ 34$\rm ^m$ 11$\rm ^s$.
Contour levels are at 1, 2, 3, 4, 5, 6, 8, 10 and 12 mJy. The lowest
(outermost) level is barely above the typical rms noise.
\end{document}